\documentclass[prl,aps,twocolumn,superscriptaddress]{revtex4-1}
\usepackage{bm}
\usepackage[colorlinks=true,linkcolor=blue,citecolor=blue]{hyperref}
\usepackage{times}
\usepackage{amsmath}
\usepackage{amssymb}
\usepackage{amsthm}
\usepackage{amsfonts}
\usepackage{enumerate}
\usepackage{latexsym}
\usepackage{ifpdf}
\usepackage{natbib}
\usepackage{psfrag}
\newcommand{\beq}{\begin{equation}}
\newcommand{\eeq}{\end{equation}}
\usepackage{graphicx}
\usepackage{makeidx}
\usepackage{color}
\usepackage{array}
\usepackage{multirow}
\usepackage{siunitx}
\usepackage{}

\begin{document}
\title{Hole doping in a negative charge transfer insulator}

\begin{abstract}
 $RE$NiO$_3$ is a negative charge transfer energy system and exhibits a temperature-driven metal-insulator transition (MIT), which is also accompanied by a bond disproportionation (BD) transition. In order to explore how hole doping affects the BD transition, we have investigated the electronic structure of  single-crystalline thin films of Nd$_{1-x}$Ca$_x$NiO$_3$ by synchrotron based experiments and  {\it ab-initio} calculations. For a small value of $x$, we find that the  doped holes are localized on one or more Ni sites around the dopant Ca$^{2+}$ ions, while the BD state for the rest of the lattice
remains intact. The effective charge transfer energy ($\Delta$) increases with Ca concentration and the formation of BD phase is not favored above a critical $x$, suppressing the insulating phase. Our present study firmly demonstrates that the appearance of BD mode is essential for the MIT and settles a long-standing debate about the role of structural distortions for the MIT of the $RE$NiO$_3$ series.
\end{abstract}

\author {Ranjan Kumar Patel}
\altaffiliation{Contributed equally}
\affiliation{Department of Physics, Indian Institute of Science, Bengaluru 560012, India}
\author{Krishnendu Patra}
\altaffiliation{Contributed equally}
\affiliation{S. N. Bose National Centre for Basic Sciences, Block-JD, Salt Lake, Kolkata-700106, India}
\author {Shashank Kumar Ojha}
\affiliation{Department of Physics, Indian Institute of Science, Bengaluru 560012, India}
\author {Siddharth Kumar}
\affiliation{Department of Physics, Indian Institute of Science, Bengaluru 560012, India}
\author{Sagar Sarkar}
\affiliation{S. N. Bose National Centre for Basic Sciences, Block-JD, Salt Lake, Kolkata-700106, India}
\author {J. W. Freeland}
\affiliation{Advanced Photon Source, Argonne National Laboratory, Argonne, Illinois 60439, USA}
\author {J. W. Kim}
\affiliation{Advanced Photon Source, Argonne National Laboratory, Argonne, Illinois 60439, USA}
\author {P. J. Ryan}
\affiliation{Advanced Photon Source, Argonne National Laboratory, Argonne, Illinois 60439, USA}
\author{Priya Mahadevan}
\email{priya.mahadevan@gmail.com }
\affiliation{S. N. Bose National Centre for Basic Sciences, Block-JD, Salt Lake, Kolkata-700106, India}
\author {S. Middey}
\email{smiddey@iisc.ac.in }
\affiliation{Department of Physics, Indian Institute of Science, Bengaluru 560012, India}
\maketitle

The topic of metal insulator transition (MIT) in transition metal (TM) compounds has remained at the forefront of condensed matter physics research ever since its' discovery as there are still some aspects that remain enigmatic~\cite{Imada:1998p1039}. While one would have thought that by now, the issue of what drives the insulating state would have been settled, one still finds new regimes that bring up the question again, in addition to the fact that there is no single parameter that is the answer. Classifying the regimes in which one finds insulating states, the early TM oxides are driven insulating by strong electron-electron interactions ($U$) at the TM site, with the band gap scaling with the strength of $U$~\cite{Zaanen:1985p418}. This is not to say that electron correlation effects play no role among the late TM compounds. Their presence is necessary, though, here the charge transfer energy ($\Delta$)  between the ligand $p$ and the TM $d$ states is the scale that decides the magnitude of the band gap. This would suggest that a negative charge transfer energy where one has an overlap of the anion $p$ and cation $d$ bands would lead to a metallic state. While these ideas made up the  Zaanen Sawatzky Allen phase diagram~\cite{Zaanen:1985p418}, a later extension that invoked the periodicity of the lattice found that even in this negative  $\Delta$ regime, strong $p$-$d$ hybridization between the anion $p$ and TM $d$ states can open up a band gap~\cite{Nimkar:1993p7355}.  In spite of this regime being identified in the theoretical phase diagram, usually some other effects  take over and  very few systems so far have been identified as belonging to this covalent insulator regime~\cite{Barman:1994p8475,Maiti:1998p1572,Matsuno:2002p193103}.

In this work, we consider  rare earth nickelates $RE$NiO$_3$  ($RE$= Pr, Nd, Sm, .. Lu, etc.)~\cite{Medarde:1997p1679,Catalan:2008p729,Middey:2016p305,Catalano:2018p046501}, which are found to belong to this negative $\Delta$ regime~\cite{Barman:1994p8475,Bisogni:2016p13017} and exhibit temperature driven simultaneous electronic and structural transitions. A magnetic transition also appears at the same temperature in  NdNiO$_3$ and PrNiO$_3$. The insulating state is associated with the development of structural distortions, leading to the presence of two structurally inequivalent Ni atoms in the unit cell~\cite{Mizokawa:2000p11263,Park:2012p156402,Johnston:2014p106404,Haule:2017p2045,Bisogni:2016p13017,Middey:2018p156801,Green:2016p195127}.  The one with
longer Ni-O bonds mimics the Ni$^{2+}$ ($d^8$: $S$=1) site, while the other with shorter Ni-O bonds mimics the Ni$^{4+}$ ($d^8\underline{L}^2$: $S$=0) site ($\underline{L}$ represents a hole on oxygen). This is referred as a bond disproportionated (BD) phase, instead of  a conventional charge disproportionated (CD) phase as there is no substantial charge difference between the two sites~\cite{Mizokawa:2000p11263,Park:2012p156402,Johnston:2014p106404,Haule:2017p2045}. It has been argued recently that the appearance of such BD phase in $RE$NiO$_3$ is intimately connected with the negative  $\Delta$~\cite{mandal2017driving}.

Suppression/disentanglement of the MIT  has been examined recently in order to understand the origin of the simultaneous  transitions of $RE$NiO$_3$~\cite{Middey:2016p305,Middey:2018p156801,Hepting:2014p227206,Stewart:2011p176401,Ojha:2019p235153,Stoica:2020disentangling,Georgescu:2019p14434}. Hole doping is another way to suppress the antiferromagnetic, insulating phase of $RE$NiO$_3$~\cite{Cheong:1994p1087,Garcia:1995p13563,Xiang:2010p032114}. Surprisingly, very little is known about the nature of these doped holes and the underlying mechanism behind the absence of the MIT. The recent discovery of superconductivity in hole doped NdNiO$_2$~\cite{Li:2019p624}, which lies at the boundary between the Mott-Hubbard and the charge transfer insulating regime~\cite{Jiang:2020p207004,Goodge:2021pe2007683118,Hepting:2020p381,Rossi:2020p00595}, further demands the necessity of understanding the consequences of hole doping on the electronic structure in these complex materials.
 In contrast to conventional semiconductors~\cite{Yamanouchi:1967p859}, the insulating state among the transition metal oxides has been found to persist even up to 10-20\%
doping~\cite{Chainani:1992p9976,Chainani:1993p15397,Katsufuji:1997p10145}.  Interestingly, the nature of the insulating state in doped compositions is less characterized. In some cases, the doped holes are localized, leading to polaronic distortions or charge ordering. However, there is no single mechanism that can be ascribed to what preserves the insulating state. In this work, we consider the hole doped nickelates which have a BD state at the undoped limit to examine the consequences of hole doping.
 Addressing these issues will not only  help us understand the hole doping in negative charge transfer systems, but also can enable us to formulate guidelines for achieving unconventional superconductivity in the non-cuprate family.

\begin{figure*}
	\vspace{-0pt}
\includegraphics[width=0.7\textwidth] {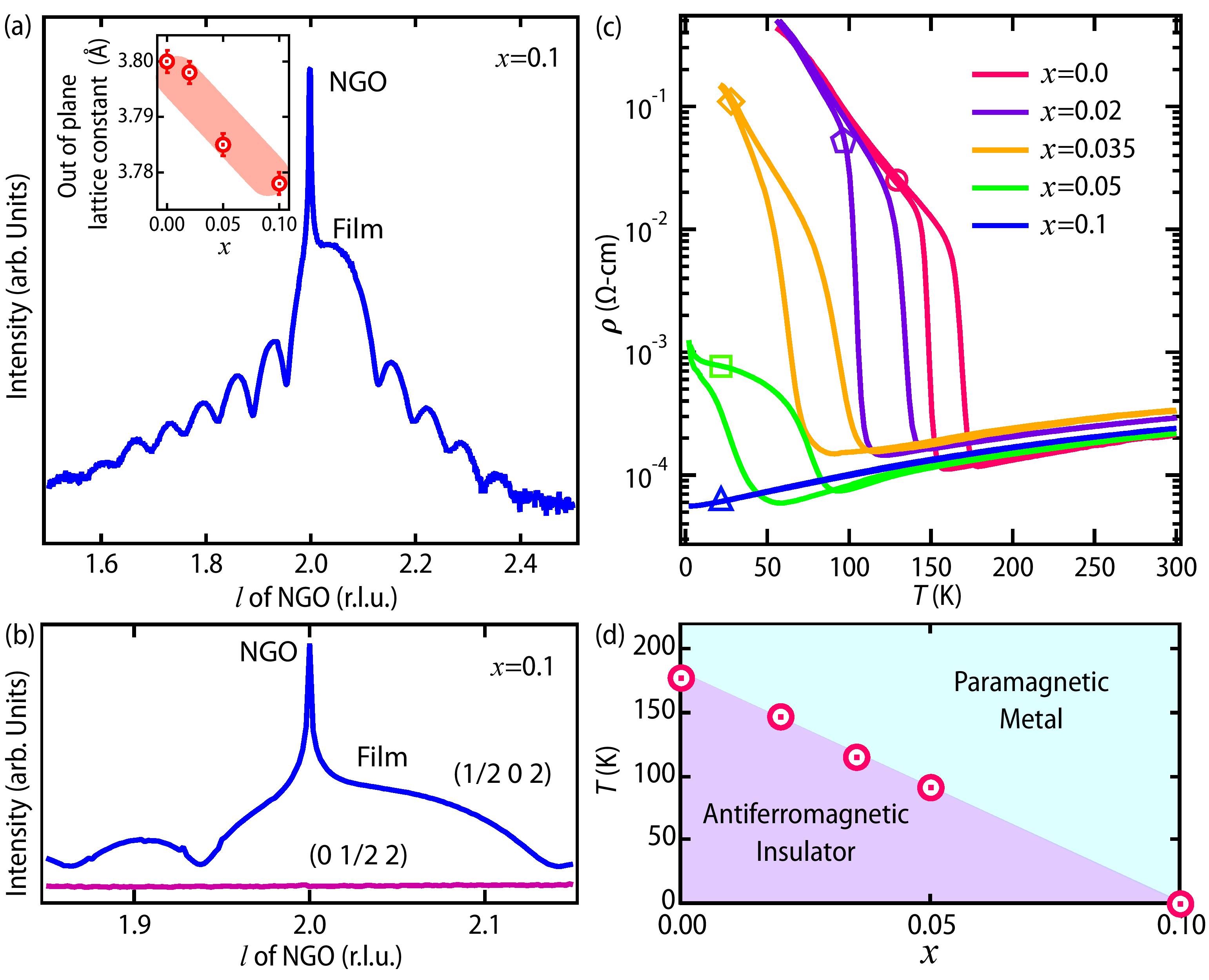}
\caption{\label{Fig1} {\textbf{Structural characterization and electronic properties. (a)} \emph{l} scan of the $x$=0.1 film through the (0 0 2)$_\mathrm{pc}$ truncation rods. The change in the out-of-plane lattice constant with hole doping has been shown in the inset.    \textbf{(b)} \emph{l} scan of $x$=0.1 sample around the (0 1/2 2)$_\mathrm{pc}$ and (1/2 0 2)$_\mathrm{pc}$ truncation rod at 300 K.   \textbf{(c)} Temperature dependence of resistivity of Ca doped NNO films.  The data for $x$=0 film has been adapted from Ref.~\onlinecite{Ranjan:2020p071601}. \textbf{(d)} Phase diagram of Nd$_{1-x}$Ca$_x$NiO$_3$ thin films. $T_\mathrm{MIT} \sim T_N$ (heating run) have been plotted as a function of $x$ (hole doping). See SM~\cite{sup} for the details about the extraction of $T_N$ from the transport data.}}
\end{figure*}

We have grown a series of single crystalline Nd$_{1-x}$Ca$_x$NiO$_3$  thin films and investigated their structural and electronic behavior by transport measurements, synchrotron based X-ray diffraction (XRD), X-ray absorption spectroscopy (XAS), and resonant  X-ray scattering (RXS).  The appearance of a simultaneous BD transition with the MIT for the parent compound NdNiO$_3$ (NNO) has been confirmed by  RXS experiments. The MIT temperature decreases linearly with the Ca doping. Our density functional theory (DFT) based calculations  carried out for small hole doping have found that instead of melting the bond ordered lattice, the doped holes are localized on one or more Ni$^{2+}$ sites.  This converts a Ni$^{2+}$ site into a Ni$^{3+}$ site, while the rest of the sites remain bond ordered. Our XAS measurements have revealed that the Ca  doping leads to an increase  in the value of $\Delta$. However, the overall value of $\Delta$ still remains negative, which preserves the BD transition. A larger amount of  Ca doping suppresses the BD transition, which has been  confirmed by our RXS measurements on the entirely metallic  film with $x$=0.1 composition.

\begin{figure*}
	\centering
		{~}\hspace*{-0cm}
	\includegraphics[scale=0.55]{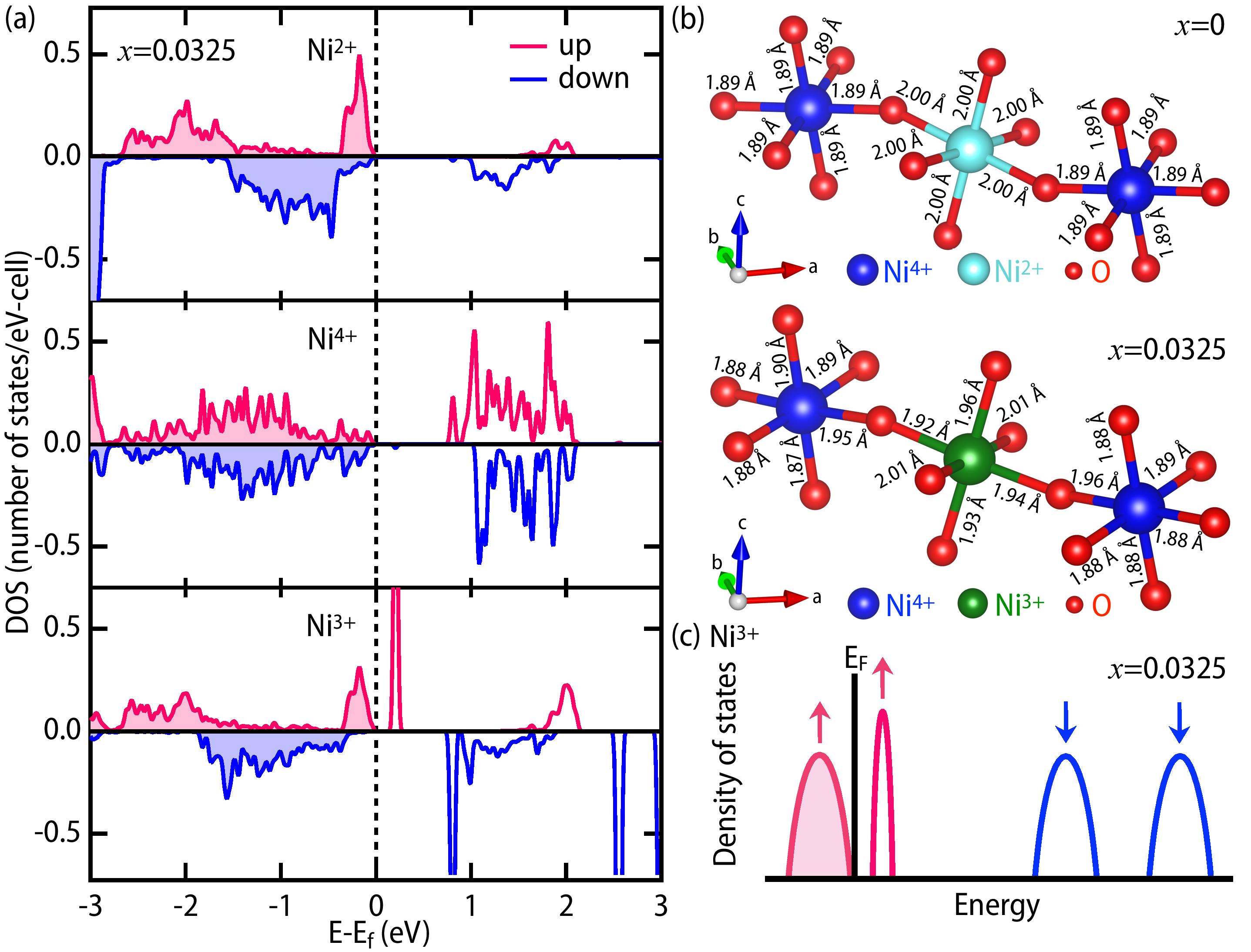}
		\caption{\label{Fig2} {\textbf {\textit{Ab initio} calculations (a)} Calculated spin resolved DOS of Ni$^{2+}$ (upper panel), Ni$^{4+}$ (middle panel), and Ni$^{3+}$ (lower panel) for $x$=0.0325. \textbf{(b)}   Ni-O bond lengths for $x$=0 and $x$=0.0325.  The doped hole is going to one of the Ni$^{2+}$ sites around the Ca dopant, converting it to Ni$^{3+}$ (see SM~\cite{sup} for  full relaxed structures). \textbf{(c)} Schematic  of DOS of Ni$^{3+}$} of $x$=0.0325.}
	\end{figure*}

\section{Results}
  15 unit cell (u.c. in pseudo-cubic setting) Nd$_{1-x}$Ca$_x$NiO$_3$ (NCNO) epitaxial thin films with $x$=0, 0.02, 0.035, 0.05 and 0.1 were grown on single crystalline  NdGaO$_3$ (1 1 0)$_\mathrm{or}$ substrates [= (0 0 1)$_\mathrm{pc}$, or and pc denote orthorhombic and pseudo-cubic setting, respectively]  by a  pulsed laser deposition (PLD)  system. Fig.~\ref{Fig1}(a) shows a representative \emph{l} scan along (0 0 2)$_{pc}$ truncation rod using synchrotron X-ray diffraction for the $x$ = 0.1 film (see Supplemental Material (SM)~\cite{sup})  for \emph{l} scans of other samples]. The diffraction patterns consist of a very sharp substrate peak, a broad film peak, and a set of thickness fringes, which arise due to the difference in the optical path of the X-ray in the film and the substrate. Surprisingly, the out-of-plane lattice constant ($c_\mathrm{pc}$) of these films decreases with the increase of Ca doping [inset of  Fig.~\ref{Fig1}(a)] though the ionic radii of Ca$^{2+}$ (1.12 \AA) is slightly greater than that of Nd$^{3+}$ (1.109 \AA).  This observation also strongly negates the presence of any oxygen vacancy (OV), which would result in an increase in the lattice volume otherwise~\cite{Nikulin:2004p775}. This finding rather implies that the doped holes are located on Ni-sites~\cite{advs.201901073,Tan:1993p12365}.

  Since  the electronic behaviors of $RE$NiO$_3$ are intimately connected with the underlying crystal symmetry, we have also investigated the structural symmetry of these films.  The parent compound NNO is orthorhombic  at room temperature. The observation of half order spots in Reflection high energy electron diffraction (RHEED) images of all of these films (see SM~\cite{sup}) suggest that these films have orthorhombic symmetry.  The orthorhombic symmetry for the $x$=0.1 film has been further confirmed by the presence [Fig.~\ref{Fig1}(b)] of  (0 1/2 2)$_\mathrm{pc}$ diffraction peak (recorded using  synchrotron X-ray), which arises due to the antiparallel displacement of $A$ sites for orthorhombic $AB$O$_3$ perovskite~\cite{Middey:2018p156801,Middey:2018p081602}. Moreover, the absence of (1/2 0 2)$_\mathrm{pc}$ peak for both film and substrate [Fig.~\ref{Fig1}(b)] ensures the film is structurally a single domain  with the same in-plane doubling direction as the substrate, which is desired for X-ray scattering experiments - described in the latter part of this paper.

All bulk $RE$NiO$_3$, having an orthorhombic phase in the metallic phase, undergo a temperature driven  MIT. The resistivity of all Nd$_{1-x}$Ca$_x$NiO$_3$ films are very similar at room temperature [Fig.~\ref{Fig1}(c)]. This further implies that the Ca doping does not incorporate OVs in these films, which would lead to an increase of resistivity in  the metallic phase~\cite{Nikulin:2004p775}.  As reported earlier~\cite{Liu:2013p2714,Mikheev:2015p1500797,Ranjan:2020p071601}, NNO film on NGO undergoes a temperature driven first order MIT with a thermal hysteresis.  The transition temperature  in both heating ($T_\mathrm{MIT}^\mathrm{h}$) and cooling run   ($T_\mathrm{MIT}^\mathrm{c}$) decreases monotonically with increasing Ca doping. Interestingly, the width of the hysteresis  ($\delta T$=$T_\mathrm{MIT}^\mathrm{h}$-$T_\mathrm{MIT}^\mathrm{c}$), which represents the coexistence of  metallic and insulating regions~\cite{Alsaqqa:2017p125132}, increases with $x$. The insulating phase is completely absent in the $x$=0.1  film.  The paramagnetic to $E^\prime$-AFM transition temperature ($T_N$) of $RE$NiO$_3$ can be approximately found out by plotting $d(ln\rho)/d(1/T)$ as a function of $T$~\cite{Ranjan:2020p041113}. A similar analysis for these films  with $x$=0.0, 0.02, 0.035, and 0.05 found $T_\mathrm{MIT}\sim T_N$ (see SM~\cite{sup}).  The plot of $T_\mathrm{MIT}^\mathrm{h}$   as a function of $x$  [Fig.~\ref{Fig1}(d)] demonstrates the suppression of $T_\mathrm{MIT}$ with a rate of 1750 K per one Ca doping within one formula unit of NNO.

\begin{figure*}
	\vspace{-0pt}
	\includegraphics[width=0.7\textwidth] {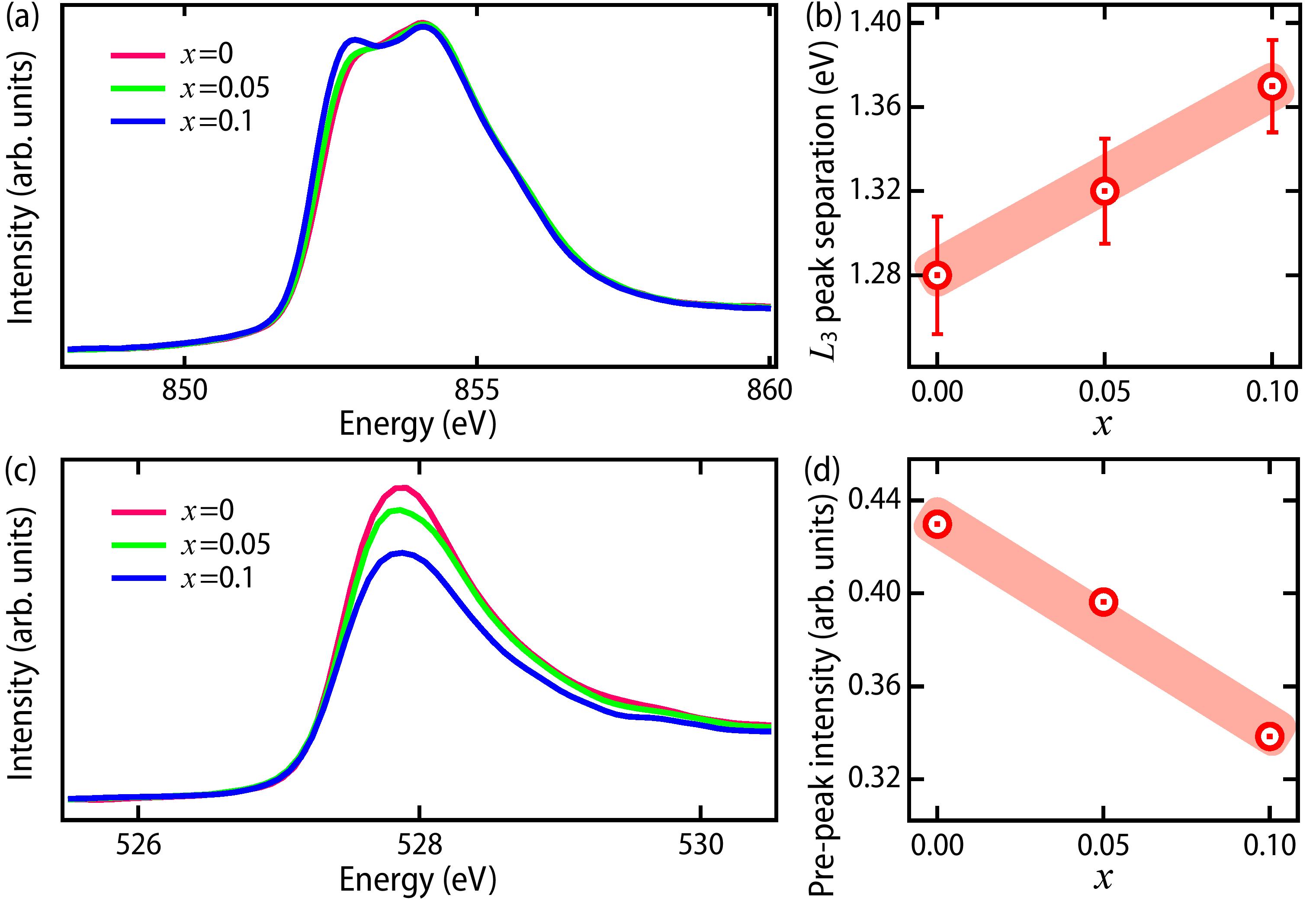}
	\caption{\label{Fig3} \textbf{X-ray absorption spectroscopy.} Normalized XAS spectra at (a) Ni $L_{3}$-edge and \textbf{(b)} the energy separation between the two peaks of the $L_3$-edge for $x$=0, 0.05, and 0.1 thin films on NGO substrate. \textbf{(c)} O $K$-edge XAS pre-peak and \textbf{(d)} the pre-peak intensity as a function of Ca doping.}
\end{figure*}

\begin{figure*}
	\vspace{-0pt}
	\includegraphics[width=0.7\textwidth] {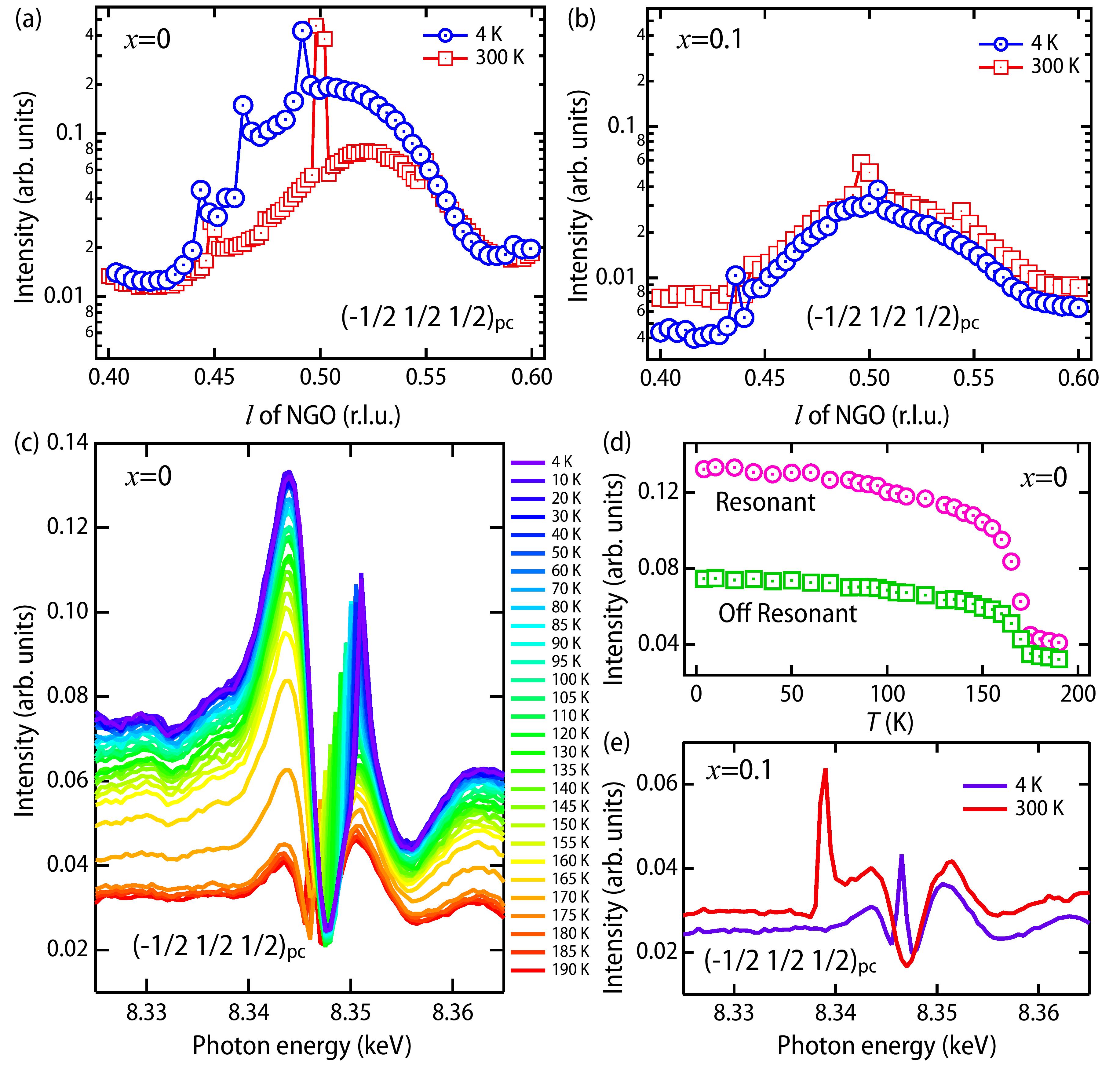}
	\caption{\label{Fig4} \textbf{Temperature dependent resonant X-ray scattering.} $l$-scan through  the (-1/2 1/2 1/2)$_\mathrm{pc}$ [(0 1 1)$_{or}$] reflection for \textbf{(a)} $x$=0 and  \textbf{(b)} $x$=0.1 thin films at 4 K and 300 K, recorded with photon energy 8.33 eV. The variation of the (-1/2 1/2 1/2)$_\mathrm{pc}$ reflection of the $x$=0 thin film around Ni $K$-edge at various temperatures for  \textbf{(c)}  $x$=0 and (e)  $x$=0.1 thin film. Temperature dependent resonant and off-resonant intensities ($I_\mathrm{res}$ and $I_\mathrm{offres}$) of the corresponding scan of $x$=0 film are plotted in \textbf{(d)}. The spikes in panels \textbf{(a)}, \textbf{(b)}, \textbf{(c)}, and \textbf{(e)} arise due to multiple waves diffraction.}
\end{figure*}

 To understand the evolution of the electronic structure with Ca doping, it is crucial to know the location of the doped hole. To examine this, we have performed {\it ab initio} calculations considering $E^\prime$-AFM state for the doping concentration of  0\%, 3.125\%, 6.25\% and 9.375\%. The system was found to be insulating for all cases. We do not attempt to find the point of the crossover between the insulating state to the metallic state in our calculations, as it is well known that the relative differences between different magnetic configurations are not correctly reproduced in DFT calculations at the undoped limit~\cite{Gou:2011p144101,Park:2012p156402,Liu:2013p2714,Hampel:2017p165130}. The trends in the structural parameters, however, have been found to be well captured which has been the basis for examining the structural and electronic properties for the doped concentrations.

 Considering the dopant concentration of 3.125\%, we find that structural distortions about the different Ni atoms lead us to 3 inequivalent types of Ni atoms. These are differentiated by the bond lengths of the NiO$_6$ octahedra. The sites identified as Ni$^{2+}$ have six long Ni-O bonds of length nearly 2.0 \AA, while those identified as Ni$^{4+}$ have six short Ni-O bonds of length nearly 1.9 \AA. The Ni$^{3+}$ sites are intermediate,
with  a combination of short and long bonds. Examining the Ni $d$ projected density of states, we find that the Ni$^{2+}$ sites have the states with $t_{2g}$ symmetry filled up in the majority as well as the minority spin channel, while the states
with $e_g$ symmetry are filled in only the majority spin channel. This is shown in Fig.~\ref{Fig2}(a) with the unoccupied minority spin $e_g$ states found in the energy window 3-4 eV above the Fermi level. The sites identified as Ni$^{4+}$ however, have the
$t_{2g}$ states in both spin channels completely filled, while the states with $e_g$ symmetry are completely unoccupied [middle panel of Fig.~\ref{Fig2}(a)].  Most importantly, some of the Ni$^{2+}$ sites around the Ca site are seen to convert to Ni$^{3+}$, with a splitting of the majority spin $e_g$ states [lower panel of Fig.~\ref{Fig2}(a)]. This is accompanied by a structural distortion [Fig.~\ref{Fig2}(b)], with four Ni-O bonds of the Ni$^{3+}$ site becoming shorter, compared to what we had for the Ni$^{2+}$ site. There are corresponding changes in the bond lengths of the neighboring Ni$^{4+}$ sites as well to accommodate these changes.  As a consequence of these distortions of the oxygens about the Ni$^{3+}$ sites,
the degenerate majority spin $e_g$ orbitals split up, with the $d_{x^2 - y^2}$ orbitals having higher energy than the d$_{3z^2-r^2}$.
This is depicted schematically in Fig.~\ref{Fig2}(c). Consequently, the doped hole is localized at the Ni$^{3+}$ sites, with the system remaining insulating~\cite{Iglesias:2021p035123}.  The sites identified as Ni$^{2+}$, Ni$^{3+}$ and Ni$^{4+}$ have the same charge and can be differentiated with only their magnetic moments (the magnetic moments of Ni$^{2+}$, Ni$^{3+}$ and Ni$^{4+}$ are 1.5 $\mu_\text{B}$, 0.98 $\mu_\text{B}$ and 0 $\mu_\text{B}$, respectively) as has been discussed earlier~\cite{Mahadevan:2001p066404}. The picture at the higher doping concentrations of 6.25\% and 9.375\% remains the same, with only more Ni$^{2+}$ sites being converted to Ni$^{3+}$,   though the distortions of the Ni$^{3+}$ sites may be different. Interestingly, such localized charge transfer has been also found in nickelate superlattice~\cite{Wang:2020p2005003}.

 The question that follows at this point is, why is the doped hole in Ca substituted NNO localized at a single Ni site, converting that to Ni$^{3+}$, leaving the bond ordering of the rest of the lattice intact. This can be understood qualitatively by the following arguments. The hole doping is accompanied by a change in the electrostatic potential, which then goes on to modify the onsite energies, thereby changing   $\Delta$~\cite{Imada:1998p1039}. When the hole is localized at the Ni$^{3+}$ site, $\Delta$ is found to change by 0.29 eV from the value of Ni$^{2+}$ site. A homogeneously distributed hole would lead to a change of 0.009 eV in the value of $\Delta$, still keeping it in the negative $\Delta$ regime. This is the reason why the bond order is not destroyed upon hole doping and survives up to 9.375\% doping as examined within our calculations.   While our calculations examine an ordered distribution of the Ca dopant sites in the lattice, these may in reality be randomly distributed. This could lead to enhanced localization than what we find here. However, the gross features of the structural distortions and the electronic structure should follow through there.    The aspects focusing on the randomness will be examined in future works.

To investigate any change in $\Delta$ with Ca doping, we have measured XAS at the Ni $L_{3,2}$-edges at 300 K (see SM~\cite{sup} for full spectra) using bulk-sensitive total fluorescence yield mode. The weak shoulder peak (around 853 eV) in the $L_3$-edge of the parent compound becomes more prominent with the Ca doping [Fig.~\ref{Fig3}(a)]. Our analysis of the $L_3$-edges~\cite{Liu:2011p161102} has also found an increase in the peak separation energy with Ca incorporation [Fig.~\ref{Fig3}(b)]. While the XAS results for the insulating phase of $RE$NiO$_3$ have been interpreted using double cluster calculations in recent times~\cite{Green:2016p195127}, single-site cluster calculations can still be used to understand the change in XAS in the metallic phase where all Ni sites are equivalent. According to the results of single-site cluster calculations~\cite{Freeland:2016p56,Liu:2011p161102,Meyers:2013p075116}, the increase in the peak separation of $L_3$ edge by 0.09 eV from $x$=0.0 to $x$=0.1 sample would correspond to approximately 0.3 eV increase of  $\Delta$. This change in $\Delta$ would decrease the covalency between Ni $d$ and O $p$ states.  This is further corroborated by  the measurement of O $K$-edge XAS (see SM~\cite{sup} for full spectra), where a pre-peak arises around 528.5 eV [Fig.~\ref{Fig3}(c)] due to the $d^8\underline{L} \rightarrow \underline{c}d^8$ transition ( $\underline{c}$ denotes a core hole in O 1$s$ core level). The intensity of the pre-peak, which is a measure of the contribution of the $d^8\underline{L}$ configuration, decreases with Ca doping in NNO [Fig.~\ref{Fig3}(d)]. We believe that the increase of $\Delta$ with Ca doping is related to the increase of (Nd, Ca)-O covalency ~\cite{Singh:2007p085102}.

Based on these experimental  and theoretical results, we propose the following situations. For a small amount of Ca doping,  the effective $\Delta$ is still negative,  and the system undergoes a BD insulating phase with the lowering of temperature~\cite{mandal2017driving}. With the increase of $x$, $\Delta$ increases and the appearance of BD phase is prohibited above a critical value. Consequently, the sample remains metallic down to low temperature. In order to confirm this, we have performed temperature dependent RXS experiments  on the $x$= 0 and 0.1 films.  Such RXS experiments around  the (0 $k$ $l$)$_\mathrm{or}$ and ($h$ 0 $l$)$_\mathrm{or}$ reflections  ($h, k, l$: odd integers) with the energy tuned to Ni $K$-edge (1$s \rightarrow$ 4$p$ transition) were employed earlier to detect CD/BD transition of several $RE$NiO$_3$ thin films and heterostructures~\cite{Staub:2002p126402,Lorenzo:2005p045128,Scagnoli:2005p155111,Lu:2016p165121,Meyers:2015p235126,Middey:2018p156801,Kim:2020p127601,Ranjan:2020p041113}. As the energy-dispersive  correction factors for the Ni sites are vanishingly small for an energy few eV away from the resonance edge~\cite{intensity}, recording of (-1/2 1/2 1/2)$_\mathrm{pc}$ reflection using X-ray with  off-resonant energy can also probe BD, which is a pure structural effect~\cite{Kim:2020p127601}. While  the  (-1/2 1/2 1/2)$_\mathrm{pc}$ reflection is forbidden in the orthorhombic phase of $RE$NiO$_3$, we observed a small intensity for this reflection  at 300 K for the NNO film [Fig.~\ref{Fig4}(a)]. This arises due to the contribution of the off-diagonal elements of the energy-dispersive scattering factors~\cite{Lu:2016p165121}. The  reflection is much stronger at 4 K, as expected for the presence of BD in the insulating phase. The intensity of this reflection does not change with lowering $T$ for the film with $x$=0.1 [Fig.~\ref{Fig4}(b)], signifying the entire suppression of the periodic BD ordering by 10\% Ca doping.

  The above conclusions  have been further validated by energy scans across the Ni K-edge for the (-1/2 1/2 1/2)$_\mathrm{pc}$ reflection.  In the metallic phase of the NNO film, we observe  some energy dependence  [Fig.~\ref{Fig4}(c)]  due to the octahedral tilt, similar to the recent reports on NNO film~\cite{Lu:2016p165121} and 1EuNiO$_3$/1LaNiO$_3$ superlattice~\cite{Middey:2018p156801}.  The increase in the off-resonant intensity  [Fig.~\ref{Fig4}(c), (d)]  across the MIT testifies a simultaneous BD transition.  Unlike the earlier reports of charge ordering in  thick NNO film~\cite{Staub:2002p126402} and  1EuNiO$_3$/1LaNiO$_3$ superlattice~\cite{Middey:2018p156801}, our 6 nm thin NNO film does not exhibit a large increase in the intensity at the resonant energy  in its insulating phase [Fig.~\ref{Fig4}(c)]. We further emphasize that the BD, which involves electronic rearrangement within the oxygen sublattice, can also drive a small electron disproportion within the Ni sublattice~\cite{Green:2016p195127,Kim:2020p127601,Ranjan:2020p041113}. The small increase of the intensity  at the resonance energy in the insulating phase  [Fig.~\ref{Fig4}(c), (d)] supports such  a scenario. In the case of $x$=0.1 film, the small resonance profile does not change across the temperature of our measurements, confirming the absence of any BD ordering [Fig.~\ref{Fig4}(e)]. These experimental results demonstrate that the $p$-$p$ gap of a covalent insulator~\cite{Nimkar:1993p7355,Barman:1994p8475} vanishes when the BD is absent and hence, the system becomes metallic. This settles a long standing issue regarding the coupling among structural effects and MIT of $RE$NiO$_3$ and further implies that the electronic and structural order parameter should be considered together to describe the simultaneous transitions of the rare-earth nickelate series~\cite{Mercy:2017p1677,Middey:2018p156801,Hampel:2019p5,Georgescu:2019p14434,Peil:2019p245127,Georgescu:2021energy}.

\section{Discussion}
To summarize, we have investigated the electronic and structural behavior of  a series of Nd$_{1-x}$Ca$_x$NiO$_3$ thin films  using advanced synchrotron measurements and DFT calculations. We have observed that  the partial replacement of Nd by Ca not only dopes holes in the system but also increases $\Delta$ gradually.  As the BD emerges for negative $\Delta$, an increase in $\Delta$ is expected to push the system away from the BD state even at small Ca doping. Interestingly, we find from DFT that the changes in $\Delta$ with respect to the undoped limit  is not the same for each Ni site. Rather than disrupting the BD state even for small $x$, we find that one or more Ni$^{2+}$ sites become Ni$^{3+}$.  This could emerge from the large structural energy cost involved in losing the BD state, in addition to the large exchange splitting that exists for the Ni$^{2+}$ sites. However, the energy cost associated with the polaronic distortion also enhances with $x$ and eventually favors delocalized doped holes above a  critical Ca doping concentration.

Finally, we would like to emphasize that the nature of hole doping in perovskite nickelate is very different compared to that in superconducting infinite-layer nickelates. While the hole doping results in an increase of $d^9\underline{L}$ state in hole doped NdNiO$_2$~\cite{Goodge:2021pe2007683118}, we have found a suppression of $d^8\underline{L}$ configuration in NdNiO$_3$, which might be responsible for the absence of the superconducting phase in the hole doped perovskite nickelates.

\section{Methods}

\noindent\textbf{Sample preparation and measurements.}
 The details of the Ca doped NNO polycrystalline target preparation and thin film growth conditions can be found in our previous work~\cite{Ranjan:2020p071601}. The growth conditions of the doped films are the same as the undoped ones. The in-situ growth was monitored using a high pressure RHEED attached with the PLD (RHEED pattern and oscillations  have been shown in SM~\cite{sup}). X-ray diffraction patterns were recorded using a  laboratory-based  diffractometer and by synchrotron X-ray diffraction. The synchrotron XRD measurements were carried out at the Indian Beamline, Photon Factory, KEK, Japan.  The electrical transport properties of these samples were measured in four probe Van der Pauw geometry in an Oxford Integra LLD system and all the contacts were made by using silver paint. XAS  and RXS measurements were carried out at  4-ID-C and 6-ID-B beam-line, respectively in Advanced Photon Source (APS), Argonne National Laboratory.

\noindent\textbf{DFT calculations.}
The DFT calculations have been carried out by imposing the non collinear $E^\prime$-type antiferromagnetic ($E^\prime$-AFM)  structure on the Ni sites in NdNiO$_3$~\cite{Scagnoli:2006p100409}.
Nd atoms have then been replaced by Ca atoms in an 160 atom supercell ($2\sqrt 2 \times 2\sqrt2 \times 2$) corresponding to the doping concentrations of 3.125\%, 6.25\%, and 9.375\%. The electronic structure was then calculated within a projected augmented wave~\cite{PhysRevB.50.17953} implementation of density functional theory within the Vienna ab-initio simulation package (VASP) code~ \cite{PhysRevB.47.558,PhysRevB.49.14251,PhysRevB.54.11169}. We have used the generalized gradient approximation (GGA)  for the exchange correlation functional~\cite{PhysRevLett.77.3865}. Electron-electron interaction effects were included on the $d$ orbitals at the Ni site, considering a value of $U$ of 4 eV within the Dudarev implementation of GGA + $U$~\cite{PhysRevB.57.1505}. The lattice parameters were fixed at the experimental values~\cite{PhysRevB.79.134432,PhysRevB.52.13563}, the internal positions were optimized till the forces were less than $10^{-3}$ eV/\AA. We used Monkhorst-Pack~\cite {PhysRevB.13.5188} $k$-points grid of $2\times2\times1$ for the calculations. A cutoff energy of 400 eV was used for the
plane wave basis states included in the calculations. The density of states and magnetic moments were calculated by considering
spheres of radii 1 \AA\ around each atom, a distance which represents almost half the Ni-O bond length. The changes in $\Delta$ on doping are quantified by changes in the electrostatic potential at the Ni and O sites.

%


\section{Acknowledgement}
RKP acknowledges XRD facility at the Department of Physics, IISc Bangalore. SM acknowledges  a SERB Early Career Research Award (ECR/2018/001512) and a  DST Nanomission grant
(Grant No. DST/NM/NS/2018/246) for funding.
PM acknowledges SERB for funding through SPF/2021/000066.
SM, JWF and PM are supported by Indo-US Joint Centre for Rational Engineering of Quantum Materials under
Indo-U.S. Science and Technology Forum. This research used resources of the Advanced Photon Source, a U.S. Department of Energy Office of Science User Facility operated by Argonne National Laboratory under Contract No. DE-AC02-06CH11357.



\section{Corresponding authors}
Srimanta Middey: smiddey@iisc.ac.in

Priya Mahadevan: priya.mahadevan@gmail.com


\end{document}